\documentclass[twocolumn]{aastex7}

\usepackage{amssymb}
\usepackage[usenames,dvipsnames,svgnames]{xcolor}
\usepackage{amsmath, mathrsfs}
\usepackage{amsfonts}
\usepackage{hyperref}
\usepackage{soul}
\usepackage[normalem]{ulem}

\definecolor{darkgreen}{rgb}{0,.5,0}
\definecolor{grey}{rgb}{.5,.5,.5}

\newfont{\nf}{cmfib8 at 10pt}

\newcommand{\add}{\textrm} 

\newcommand{\mj}{\,$\mathrm{M_J}$}
\newcommand{\au}{\,au}
\newcommand{\um}{\,$\mu$m}

\newcommand{\Lsolar}{\,L$_{\odot}$}

\newcommand{\uJy}{\,$\mu$Jy}

\newcommand{\wdtwenty}{WD\,2105$-$82}
\newcommand{\wdtwelve}{WD\,1202$-$232}
\newcommand{\epochtwenty}{2023~April~21}
\newcommand{\epochtwelve}{2023~February~9}
\newcommand{\epochB}{2025~June~3}

\begin{document}
\title{Follow-up Observations of Candidate White Dwarf Planets with MIRI}
\shorttitle{No, non si muove}
\shortauthors{Mullally et al.}

\author[0009-0004-7656-2402]{Fergal Mullally}
\affiliation{Constellation, 1310 Point Street, Baltimore, MD 21231}
\email{fergal.mullally@gmail.com}

\author[0000-0001-7106-4683]{Susan E. Mullally}
\affiliation{Space Telescope Science Institute, 3700 San Martin Dr, Baltimore, MD 21218, USA}
\email{smullally@stsci.edu}

\author[0000-0002-7698-3002]{Misty Cracraft}
\affiliation{Space Telescope Science Institute, 3700 San Martin Dr, Baltimore, MD 21218, USA}
\email{cracraft@stsci.edu}

\author[0000-0001-9729-9413]{Samantha N. Bianco}
\affiliation{Space Telescope Science Institute, 3700 San Martin Dr, Baltimore, MD 21218, USA}
\email{sbianco@stsci.edu}

\author[0000-0003-0475-9375]{Lo{\"i}c Albert}
\affiliation{Institut Trottier de recherche sur les exoplanètes and Département de Physique, Université de Montréal, 1375 Avenue Thérèse-Lavoie-Roux, Montréal, QC, H2V 0B3, Canada}
\email{loic.albert@umontreal.ca}

\author[0000-0002-1783-8817]{John Debes}
\affiliation{AURA for ESA, Space Telescope Science Institute, 3700 San Martin Dr, Baltimore, MD 21218, USA}
\email{debes@stsci.edu}

\author[0000-0001-5941-2286]{J. J. Hermes}
\affiliation{Department of Astronomy, Boston University, 725 Commonwealth Avenue, Boston, MA 02215, USA}
\email{hermes@bu.edu}

\author[0000-0001-6098-2235]{Mukremin Kilic}
\affiliation{Homer L. Dodge Department of Physics and Astronomy, University of Oklahoma, 440 W. Brooks St, Norman, OK 73019, USA}
\email{kilic@ou.edu}


\author[0000-0001-8362-4094]{William T. Reach}
\affil{Space Science Institute, 4765 Walnut Street, Suite 205, Boulder, CO 80301, USA}
\email{wreach@spacesciece.org}




\begin{abstract}
We report on second-epoch imaging of two candidate planet-hosting white dwarfs stars, \wdtwenty\ and \wdtwelve. Both stars showed evidence of resolved, planet-mass candidate companions in observations using the MIRI mid-infrared imager on JWST. WD2105 also showed evidence of an infrared excess consistent with an unresolved 1.4\mj\ companion with an orbital separation of $<$4\au. Our second epoch observations confirm that the source of the excess shares common proper motion with the star, 
\add{ The excess is almost certainly due to a companion planet or debris disk}. However, neither of the two resolved sources with projected separations of $>1$\,arcsec in the first epoch of JWST observations show measurable proper motion and are thus likely  faint, unresolved background galaxies. \add{We also search for common proper motion companions out to hundreds of \au, but find no evidence of widely separated companions.}
\end{abstract}


\section{Introduction}
The presence of metals like calcium and iron in the photospheres of white dwarf (WD) stars is a long standing puzzle. The strong gravitational field pulls heavier elements down out of the atmosphere on timescales as short as a few thousand years \citep{Koester09}, leaving a chemically pure exterior of hydrogen. 
But \citet{Zuckerman03} and \citet{Koester14} showed that at least 25\% of isolated, hydrogen-atmosphere WDs (the DA spectral class) show metal-lines in their spectra (forming the DAZ subclass), and recent analyses suggest that at least 38\% of DA are
actively accreting metals from an external source \citep{OuldRouis24}. 

The widely adopted hypothesis is that relic solar systems
are the source of the accreted material \citep{Alcock86, Jura03}. In this scenario, massive planets on long-period orbits survive the red-giant phase and occasionally perturb the orbits of asteroids and comets, which then fall in towards the WD. When these bodies pass inside the star's Roche limit they disintegrate into a cloud of dust and gas, which then accretes onto the star \citep{vonHippel07dazd}. This hypothesis gained support with the discovery of more than 50 metal-polluted WDs with either hot dust \citep[e.g.][]{Reach05b, Farihi08dust, Becklin05, Kilic05, Gentile2021, Melis2020},  and/or gas \citep[e.g.][]{Gansicke06} within the Roche limit of the WD. The direct detection of transiting disintegrating planetesimals at WD 1145+017 \citep{Vanderburg15} and several similar systems \citep{Vanderbosch20,Farihi22} confirms that tidal disruption of minor bodies is ongoing.

Using numerical simulations, \citet{DebesWalsh2012} showed that a single giant planet can perturb planetesimals into highly eccentric orbits that can create a steady stream of WD crossing planetesimals. If this giant planet theory for WD contamination is correct, every DAZ hosts a planetary system containing at least one giant planet. We can then use DAZs as independent evidence that planetary systems analogous to our own, with giant planets in long-period orbits, exist around more than one third of main-sequence stars.

\add{Conversely, \citet{Jenkins24} argue that Jupiter analogs likely do not play a significant role in driving white dwarf pollution. They found no correlation between metal pollution and white dwarf progenitor metallicity (measured by studying widely separated main-sequence binary companions). The frequency of radial velocity planets is known to scale with the square of host metallicity above the solar abundance \citep{Fischer05}. If giant planets were the primary cause of WD pollution they should be more common around remnants of metal rich stars, and we should expect these WDs to experience more pollution.}

Despite searches using a variety of techniques, few confirmed exoplanets have been found orbiting WDs.
The best characterized is the transiting giant exoplanet WD\, 1856+534~b \citep{Vanderburg2020Nature}. Recent observations by \citet{Limbach25} detected unresolved thermal emission from this exoplanet  using JWST's Mid-Infrared Instrument (MIRI); at $<$200\,K, it is the coldest exoplanet with observed thermal emission. Two separate micro-lensing events from distant, undetected sources that are presumed to be WD stars have also been reported \citep{Blackman21, Zhang24}.

WDs are intrinsically faint ($\sim 10^{-4}$\Lsolar). 
The contrast ratio between a $12{,}000$\,K WD and a Jovian mass planet at 300\,K is only $10^{-1}$ at mid-infrared wavelengths ($15-20$\um). Jovian companions to WDs at moderate orbital separations ($\gtrsim 10$\au) can be directly detected by JWST with MIRI photometry without using the coronagraph. The lack of a coronagraph, and its narrow band filters, allows for a deeper search for fainter, lower-mass companions than would otherwise be possible \citep{Poulsen2023}. Companions at closer orbital separations can be detected with lower sensitivity as excess infrared flux relative to that expected from the WD alone \citep{Livio92ir}.

\citet{Mullally24} reported on  a small JWST program (GO\,1911) to observe a sample of four DAZs with MIRI. They found two candidate planets, spatially resolved from their WDs (\wdtwelve\ and \wdtwenty), with the right brightness and spectral energy distribution and projected orbital separations of 35\au\ and 11\au. While they calculated a probability of less than one in three thousand that both detections were false alarms \citep[using galaxy counts from][]{WuCossas23}, they cautioned that the surface density of faint, unresolved galaxies at these magnitudes and wavelengths was poorly understood and labeled the sources as ``candidate planets'' until a second-epoch image could demonstrate whether the companions shared common proper motions with their putative hosts.

A re-analysis of the data by \citet{Debes25} detected excess flux from \wdtwenty\ at 15 and 21\um\ with 2.0 and 5.6\,$\sigma$ significance, and a 5$\sigma$ excess at 21\um\ around another object, WD 2149+021. They struggled to fit the data with a dust model and tentatively concluded that the excess was most likely due to a planet.

In this work, we report on second-epoch follow-up observations of these two spatially resolved  candidates with 15\um\ MIRI imagery. We confirm that the IR excess around \wdtwenty\ shares common proper motion with the star, ruling out a line-of-sight background galaxy as a false-positive source. However, the faint, resolved sources do not share common proper motion, and are indeed background galaxies.

  \begin{figure}
     \begin{center}
    \hspace*{-1cm}\includegraphics[angle=0, scale=.45]{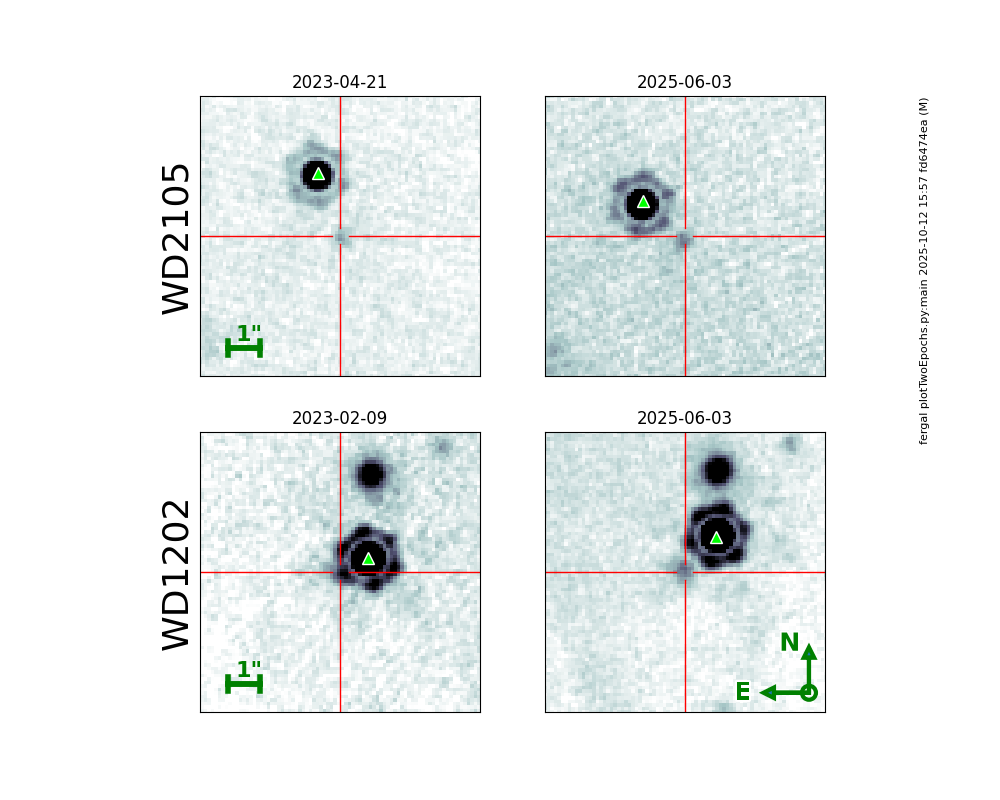}
    \caption{Discovery (left) and follow-up (right) images of \wdtwenty\ (top) and \wdtwelve\ (bottom). The images are centered on the location of the candidate companions, which are marked with red cross-hairs. The WDs, marked with a green triangle in each image, show significant proper motion relative to the candidates demonstrating that they are not physically bound. The other stationary objects are all background sources.
\label{propermotion}}
     \end{center}
 \end{figure}

\section{Observations\label{observations}}
WD\,2105$-$82 and WD\,1202$-$232 were first observed with the MIRI imager on JWST in Cycle\,1 on \epochtwenty\ and \epochtwelve\ as part of GO\,1911. These observations are described in \citet{Mullally24}. Briefly, the targets were observed at multiple wavelengths, including $12{,}088$\,s and $8414$\,s exposures at 15\um\ respectively. We show a table of observations in Table~\ref{journal}.

\begin{table}[tb]
    \centering
    \begin{tabular}{lrr}

 ~ & WD\,1202 & WD\,2105  \\

 {\bf Epoch 1}  & & \\
 Date (UTC) & 2023 Feb 9 & 2023 April 21  \\
 Bands (\um) & 5.6, 7.7, 15, 21& 5.6, 7.7, 15, 21 \\
 15\um\ Exposure (sec)& $8414$ & $12{,}088$ \\
\\
{\bf Epoch 2}  & & \\
 Date (UTC) & 2025 Jun 03 & 2025 Jun 03 \\
 Bands (\um) & 15 & 15 \\
 15\um\ Exposure (sec)& 7434 & 4304 \\

    \end{tabular}
    \caption{Journal of JWST/MIRI Observations \label{journal}}
    
\end{table}


In GO~4587, we re-observed both systems on \epochB\ as part of JWST's Cycle~3 program. \wdtwelve\ was observed with the F1500W filter on MIRI. There were 19 dither positions, with 70 groups and two integrations at each dither position, taken with the FASTR1 mode. \wdtwenty\ used an 11 point dither pattern, also using F1500W, and 70 groups with two integrations of FASTR1 mode.

 Each dataset was processed through the JWST pipeline, version 1.18.0, with the CRDS context of jwst\_1364.pmap. For the first stage of the pipeline, calwebb\_detector1, the rejection\_threshold of the jump step was set to 5.0, and the jump parameter find\_showers was set to False, in order to skip cosmic ray shower finding. Default values were used for all parameters in the pipeline's second stage, calwebb\_image2. 
 The output files were stacked, and a median ``sky background'' image was created, which was then subtracted from each of the individual files to remove the background and any remaining detector artifacts. The final stage of the pipeline, calwebb\_image3, was then run, using the sky-subtracted files as input. Most of the parameters in this pipeline used the default values, except for: in the resample step, the kernel was set to ‘square’ and the weight\_type was set to ‘exptime’, and in the outlier\_detection step, the scale values were set to ‘1.0 and 0.8’. These are the default values set in the code and parameter reference files at the time the data was processed, but were not necessarily the default in previous versions of the pipeline, so are set specifically to match the parameters used for the processing of data in GO\,1911. The calwebb\_image3 pipeline then outputs a single dither-combined, sky subtracted, resampled i2d image.

\section{Results}
\subsection{Candidate Resolved Companions}
In Figure~\ref{propermotion}, we show two epochs of 15\um\ MIRI images separated by $\approx 27$ months for both stars. The crosshairs point to the candidate planets, and the proper motions of the WDs (indicated by green triangles) are visually obvious. In each case, the candidates do not share common proper motion with the WD. 

The projected orbital motion vector of a planet can sometimes align anti-parallel with the star's proper motion vector. In certain circumstances the two vectors could conceivably cancel out, making the planet momentarily hover in the sky. However, the maximum projected orbital-motion of these candidates in face-on circular orbits will be less than a tenth of a pixel, while the WD proper motion is of order 10 pixels.  We therefore conclude that the candidates are background galaxies, and not planets physically associated with the star. 



\subsection{Confirmation of Unresolved Infrared Excess}
\add{
    \citet{Debes25}  measured a flux of 70.79\uJy\ at 15\um\ for \wdtwenty\ in Cycle 1. Assuming an absolute flux calibration uncertainty of 2\%, they calculated a 2$\sigma$ excess flux over that expected in this passband based on fitting atmosphere models from  \citet{Holberg06} to photometry at shorter wavelengths. They also found a $5 \sigma$ excess at 21\um. 
    In Cycle 3 we used the same technique as Debes to measure an almost identical flux at 15\um\ of 
    70.81\uJy. The difference in flux is many times smaller than even the photometric uncertainty of the observations, $\approx$0.2\uJy.
}

 The probability of observing a spurious $2\sigma$ excess in two separate epochs is less than 0.25\%. We therefore confirm the statistical significance of the excess and determine that the source shares the star's common proper motion and is physically associated with it. While Debes concludes that the spectral energy distribution of the excess is more consistent with a 1.4\mj\ planet than with a dust disk, our observations do not disambiguate between these two scenarios. The WD's strong magnetic field \citep[9.2 kG,][]{landstreet12} may also produce an excess through an as-yet unknown mechanism. 
We measure a flux for \wdtwelve\ of 
182.34(41)\uJy, 
consistent with both the Cycle 1 measured flux of 183.21(67)\uJy\ 
\add{(where the quoted uncertainties are based on photometric scatter alone)}
and the expected flux based on the atmosphere model.

\begin{figure}
    \begin{center}
\includegraphics[angle=0, scale=.5]{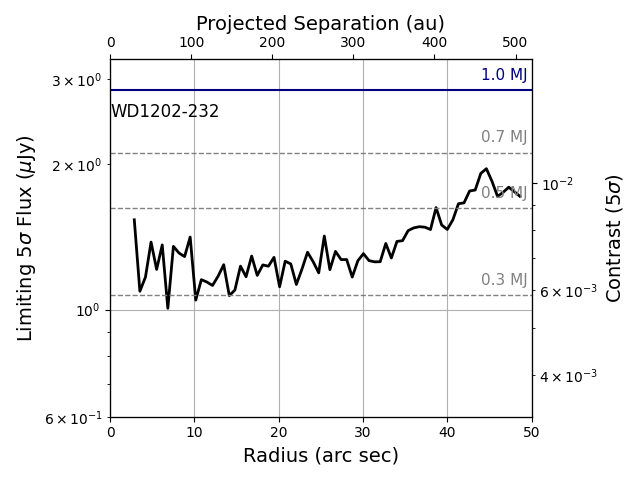}
\includegraphics[angle=0, scale=.5]{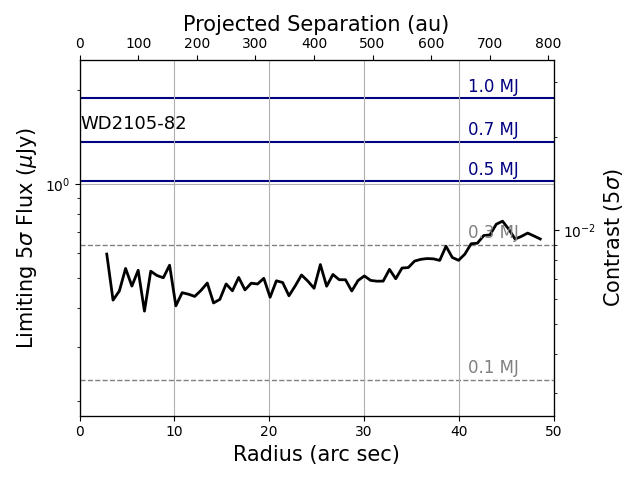}
\caption{{\bf Top:} 5$\sigma$ detection limits for resolved companions around \wdtwelve\ in units of limiting flux and contrast. The solid blue line indicates expected flux from 1.0\mj\ 6.1\,Gyr model from \citep{Linder19}. The grey dashed lines show expected flux extrapolated beyond the lower limit of the model grid.
{\bf Bottom:} Detection limits for \wdtwenty. Mass limits assume an age of 2.4\,Gyr. \label{detectionlimits}}. 

    \end{center}
\end{figure}

\subsection{Common Proper-Motion Search for Widely-Separated Companions}
Planets have been found at very wide orbits ($>$100\au) around A stars (the progenitors of most WDs) using direct imaging \citep{Nielsen2019GPI}.
Searching for very widely separated planets is challenging with single epoch MIRI imaging. The high surface density of faint, unresolved, red, background-galaxies (with spectral energy distributions similar to cool planets) means many such objects appear in every MIRI field. Single-epoch observations must restrict themselves to a small radius around the WD where the probability of chance alignment is small \citep{Mullally24}. However, background galaxies have negligible proper motion, so a search based on two epochs has no such restriction.

We co-add our images to increase our sensitivity to faint common-proper-motion companions. Using the sources from the JWST pipeline detection catalog as fiducial points, we calculate an affine transformation to align the first epoch to the coordinates of the second epoch. We then create difference images A-B and B-A between the aligned first epoch (A) and the second epoch (B) to remove all stationary stars. Companions that share proper motion with the star are not removed as they show multi-pixel shifts between both epochs. We then average the difference images after shifting one by the observed proper motion of the WD. Any object with common proper motion should be combined, enhancing the signal to noise of the detection. The shift was measured by centering the ($x$,$y$) position of the WD on both images using a 2D Gaussian. We inspect all 5$\sigma$ outliers found in this technique and determined that all were due incomplete subtraction of diffraction spikes. 


We use the method described in \citet{Poulsen2023} to estimate our detection sensitivity in the combined images, and present our results in Figure~\ref{detectionlimits}. We draw multiple 3.55\,pixel radius apertures (65\% encircled energy) in concentric circles out to 450\,pixels around the target WD. We measure the flux in each aperture, then measure the mean flux and the standard deviation in each ring after removing values with a median absolute deviation more than 5$\sigma$ from the other measurements in that ring. For \wdtwelve\ we measure a 5$\sigma$ mean standard deviation of 0.45\uJy\ and for \wdtwenty\ we get a 5$\sigma$ median standard deviation of 0.34 \uJy\ for separations between 7 and 35 arc seconds. The standard deviation in the background increases by 18\% between 35 and 45 arc seconds, likely because fewer dithers cover that part of the sky.


We used the BEX models \citep{Linder19} to convert our 5$\sigma$ detection limits into limiting masses. We assumed solar metalicity and ages comparable to the age of our WDs \citep[from][]{Debes25}. We then scaled the flux by the distance to the WD. Our detection limits are below the low-mass limit of our model grid, so we extrapolated  by fitting a straight line to the grid of log mass and log flux.  
While these extrapolated masses should be treated with considerable caution, the limits are consistent with Saturn-mass planets. 
In Figure 2, the predicted fluxes from models of different masses are indicated by dark blue lines, while extrapolated masses are represented by grey dashed lines.

We also compare the measured background flux of the two epochs separately using the concentric rings technique described above. For \wdtwelve\ we obtain the same background flux in both epochs despite the GO\,4857 exposure time being 1000\,s (12\%) shorter than  GO\,1911.  We attribute the difference to our different dither patterns. GO\,4857 has 19 dithers with 2 integrations per exposure instead of only 8 dithers with 4 integrations per exposure.

\section{Discussion}

\begin{figure*}
    \begin{center}
\includegraphics[angle=0, scale=.7]{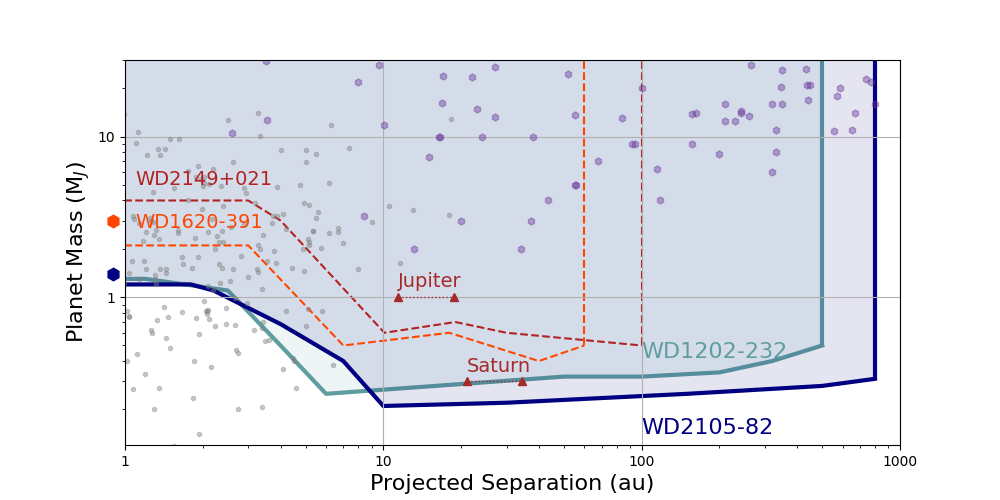}
\caption{Schematic of the detection limits for the four stars in our survey. Details of varying detection limits with orbital separation shown in Figure~\ref{detectionlimits} have been simplified, and detection limits below 0.3\mj\ are extrapolated beyond the limits of the available model grid (see text for details). Solid lines (\wdtwenty\ and \wdtwelve) indicate the two stars in this work with two epochs of data, while the dashed lines indicate two stars with a single epoch \citep[from][]{Mullally24}. The hexagons to the left of the plot indicate the inferred masses of the unresolved candidate companions. The small grey points are known radial-velocity planets, while the larger purple points indicate directly imaged planets around young A-stars from \citet{Nielsen2019GPI}. The triangles indicate the locations where Jupiter and Saturn would appear after adiabatic outward migration of \wdtwelve\ (expansion factor of 2.2, the left triangle) and \wdtwenty\ (expansion factor of 3.6, right triangle) during the mass-loss phase of their parent stars. \label{masslimits}}
    \end{center}
\end{figure*}

\subsection{Survey detection limits}
In Figure~\ref{masslimits} we show the updated detection limits for the four stars observed in this survey, combining our limits on unresolved companions within 2--3\au\ \citep{Mullally24}, partially resolved companions using the Kernel Phase Imaging search of \citet{Debes25}, and limits on resolved companions (also from Debes) updated for the two stars in this work. The hexagons indicate the masses of planets consistent with the age and observed infrared excess flux associated with \wdtwenty\ and WD\,1620 from Debes. Again, these infrared excesses have not been conclusively shown to be planets, but are included for completeness. 
Our mass limits for undetected resolved companions, while considerably uncertain, are substantially below the most sensitive white-dwarf planet search made by {\it Spitzer} \citep[5\mj\ for the WD GD\,66,][]{Mullally09}, and comparable to the most sensitive search for long-period planets around any WD \citep[0.1\mj\ at 10\au\ around G117-B15A, using the pulsation timing technique of][]{Kepler05a}. These limits highlight the power of direct-imaging with MIRI and JWST to probe for analogs of our outer solar system.

\subsection{Implications for DAZ accretion models}

\add{
Our survey is sensitive to spatially-resolved planets $> 0.5$\mj\ and 10--60\au\ around all 4 stars, but we find no planets in this region of parameter space. Binomial statistics suggest that the probability of finding $n$ planets around $N$ stars assuming an occurrence rate $f$ is given by $P(x | n, f) = {n \choose x}  f^x (1-f)^{n-x}$. If we take, $n=0$ and $N=4$ then the 2$\sigma$ ($p$=0.05) upper bound on the occurrence rate of planets in this part of parameter space is 53\%. Even with this small sample we can conclude that such widely separated Jovian analogs are not responsible for metal accretion in every star, and are most likely a minority contributor at best.
}

\add{
Inside of 10\au\, we find two candidate unresolved-companions. If we assume both candidates are indeed planets, a similar argument suggests a 2$\sigma$ upper bound on such closer-in companions is 90\%. Given the different mass sensitivities for the different target stars (with minimum detectable masses ranging from 1--4\mj\ as shown in Figure~\ref{masslimits}), our two detections would be entirely consistent with the hypothesis that every WD hosts a Jovian planet inside 10\au. MIRI medium-resolution spectroscopy could disambiguate between planets or dust for these infrared-excess systems. 
}

\add{
If future works determines that these candidates are not planets, this argues against giant planets anywhere inside 60\au\ being the drivers of accretion. If planets are the drivers they must lie beyond 60\au\ or have masses less than 1\mj.} The next obvious step to test the connection is to observe a population of DAZs to MIRI's background floor. According to the ETC such a survey will be sensitive to planets as small as Neptune (0.1\mj) for $\sim$ 10 stars. \add{If each star is observed with two epochs such a survey would also be sensitive to companions out to $\sim$1000\au.}

But perturbers might be smaller yet.
\citet{Veras23} show that planets with masses just a few times the mass of the moon can drive accretion if fortuitously placed near dense asteroid belts. However, \citet{Childs2019} show that accretion driving efficiency declines with mass of the perturber. For tiny planets to effectively drive accretion they must be either plentiful, or conveniently located relative to their asteroid or comet fields.

\section{Conclusion}
We report on a second epoch of MIRI imaging of two candidate  planetary systems around metal rich WD stars. We determine that candidate resolved companions are in fact background objects unassociated with the star. We confirm an infrared excess around one candidate (\wdtwenty) is physically associated with this star.

\begin{acknowledgements}

This work is based on observations made with the NASA/ESA/CSA James Webb Space Telescope. The data were obtained from the Mikulski Archive for Space Telescopes at the Space Telescope Science Institute, which is operated by the Association of Universities for Research in Astronomy, Inc., under NASA contract NAS 5-03127 for JWST. \add{The specific observations analyzed can be accessed via \dataset[doi: 10.17909/s9wx-3e62]{https://doi.org/10.17909/s9wx-3e62}.} Support for program numbers GO 1911 and GO 4587 was provided through grants from the Space Telescope Science Institute. MK gratefully acknowledges support from the NSF under grant  AST-2508429, and the NASA under grants 80NSSC22K0479, 80NSSC24K0380, and 80NSSC24K043. \add{We thank an anonymous referee for their comments that substantially improved this manuscript.}
\end{acknowledgements}

\facility{JWST}
\facility{MAST}
\software{ Numpy \citep{numpy}
Astropy \citep{astropy},
Matplotlib \citep{matplotlib},
Jdaviz \citep{jdaviz}
}

\bibliography{mull, dustywd}{}
\bibliographystyle{aasjournalv7}

\end{document}